\documentclass[12pt]{article}
\def\eq#1\en{\begin{equation} #1 \end{equation}}

\usepackage{graphicx,color}

\def\pp#1{\partial_{#1}}

\begin{document}

%%%%%%%%%%% CERN Titlepage %%%%%%%%%%%

\begin{titlepage}
\begin{flushright}
\begin{tabular}{l}
CERN-TH/2002-303\\
IUB-TH/021\\
hep-ph/0212292\\
November 2002 (v1)\\
August 2003 (revised)
\end{tabular}
\end{flushright}

\vspace*{0.5truecm}

\begin{center}
\boldmath
{\Large \bf The photon-neutrino interaction in
non-commutative gauge field theory and astrophysical bounds}
\unboldmath

\vspace*{0.5cm}

\smallskip
\begin{center}
{\sc { P. Schupp$^1$, J. Trampeti\'{c}$^{2,3}$, J. Wess$^{4,5}$ and G. Raffelt$^4$}}\\

\vspace*{1cm}

{\sl $^1${International University Bremen, 
School of Engeneering and Science, Campus Ring 1,
28759 Bremen, Germany}},\\
{\sl $^2${Theory Division, CERN, CH-1211 Geneva 23, Switzerland},\\
$^3${Theoretical Physics Division, Rudjer Bo\v skovi\' c Institute, 
Zagreb, Croatia}},\\
{\sl $^4${Max-Planck-Institut f\"ur Physik, F\"ohringer Ring 6, \\
80805 Munich, Germany}}, \\ 
{\sl $^5${Theoretische Physik, Universit\"at M\"unchen, Theresienstr. 37, \\
80333 Munich, Germany}}
\end{center}
\vspace{0.5truecm}
%\Sl{D} \Sl{=}
\vspace{1.5truecm}

{\large\bf Abstract\\[10pt]} \parbox[t]{\textwidth}{
In this letter we propose a possible mechanism of
left- and right-handed neutrino
couplings to photons, which arises quite naturally in non-commutative
gauge field theory.
We estimate the predicted additional energy-loss in stars induced by
space-time non-commutativity.
The usual requirement that any
new energy-loss mechanism in globular stellar clusters
should not excessively exceed the standard neutrino losses
implies a scale of non-commutative gauge theory above the
scale of weak interactions.%
%produces a reasonable lower bound on the scale of non-commutative
%gauge field theories i.e. approximately $\Lambda_{\rm NC} > 81~{\rm GeV}$.
}

\end{center}

\end{titlepage}

\thispagestyle{empty}
\vbox{}

\setcounter{page}{1}

%%% end CERN title page %%%%%%%%%%%%%

\noindent
Neutrinos do not carry a U(1)
(electromagnetic) charge and hence do not directly couple to Abelian
gauge bosons (photons) -- at least not in a commutative setting. In the presence of space-time
non-commutativity, it is, however, possible to couple neutral particles to 
gauge bosons via a star commutator. The relevant covariant derivative is
\begin{equation}
\widehat D_\mu \widehat \psi = \partial_\mu \widehat \psi - i \kappa e \widehat A_\mu \star \widehat \psi
+ i \kappa e \widehat\psi \star \widehat A_\mu \; , \label{ncc}
\end{equation}
with the $\star$--product and 
a coupling constant $\kappa e$ that corresponds to a multiple (or fraction) $\kappa$ of the
positron charge $e$.  The $\star$--product  is associative 
but, in general, not commutative -- otherwise the proposed coupling to the non-commutative
photon field $\widehat A_\mu$ would of course be zero.
In (\ref{ncc}), one may think of the non-commutative
neutrino field $\widehat \psi$ as having left charge $+\kappa e$, right charge $-\kappa e$
and total charge zero. From the perspective of non-Abelian gauge theory,
one could also say that the neutrino field is charged in a non-commutative
analogue of the adjoint representation with the matrix multiplication replaced by the $\star$--product. 
From a geometric point of view, photons
do not directly couple to the ``bare'' commutative neutrino fields,
but rather modify the non-commutative background.
The neutrinos propagate in that background.

Kinematically, a decay of photons into neutrinos is, of course, allowed only
for off-shell photons. This is
still true in a constant or sufficiently slowly varying non-commutative background:
Such a background does not lead to a violation of four-momentum conservation,
although it may break other Lorentz symmetries. 

Physically, such a coupling of 
neutral particles to gauge bosons is possible because the non-commutative
background is described by an antisymmetric tensor $\theta^{\mu\nu}$
that plays the role of an external field in the theory \cite{sny}--\cite{ws1}. 
The $\star$--product in (\ref{ncc}) is a (non-local) bilinear expression in
the fields and their derivatives that takes the form of a series in $\theta^{\mu\nu}$. 
To lowest order we obtain
\[
\widehat D_\mu \widehat \psi = \partial_\mu \widehat \psi +  
\kappa e \theta^{\nu\rho} \, \partial_\nu\widehat A_\mu \, \partial_\rho \widehat \psi
\; 
.
\]
A similar expansion (Seiberg-Witten map) exists 
for the non-commutative fields $\widehat \psi$, $\widehat A_\mu$
in terms of $\theta^{\mu\nu}$, ordinary `commutative' fields $\psi$, $A_\mu$ and their derivatives.
The scale of non-commutativity $\Lambda_{\rm NC}$ is fixed by choosing dimensionless matrix elements
$c^{\mu\nu}=\Lambda^2_{\rm NC} \theta^{\mu\nu}$ of order one.
Gauge invariance requires that all $e$'s in the action should be multiplied by $\kappa$. To the order
considered in this letter, $\kappa$ can be absorbed in a rescaling of $\theta$, i.e. a rescaling of
the definition of $\Lambda_{\rm NC}$.

The coupling (\ref{ncc}) is part of an effective model of 
particle physics involving neutrinos and photons
on non-commutative space-time. It
describes the scattering of particles 
that enter from an asymptotically commutative
region into a non-commutative interaction region.
The model satisfies the following requirements \cite{sny}--\cite{ws1}:
\begin{enumerate}
\item[(i)] Non-commutative effects are described perturbatively. The action is
written in terms of asymptotic commutative fields. 
\item[(ii)] The action is gauge-invariant under U(1)-gauge transformations.
\item[(iii)] It is possible to extend the model to a non-commutative electroweak
model based on the gauge group U(1)$\times$SU(2). An appropriate non-commutative 
electroweak model with $\kappa =1$ 
can in fact be constructed with the same tools that were used for
the noncommutative standard model of \cite{calmet}.
\footnote{For a model in which only the neutrino has dual left and right charges,
$\kappa = 1$ is required by the gauge invariance of the action.}
\end{enumerate}

The action of such an effective model
differs from the commutative theory essentially by the presence of star products
and Seiberg--Witten (SW) maps. The Seiberg--Witten maps~\cite{SW} are necessary to express
the non-commutative fields $\widehat \psi$, $\widehat A_\mu$ that appear in the action
and transform under non-commutative gauge transformations, in terms
of their asymptotic commutative counterparts $\psi$ and $A_\mu$.
The coupling of matter fields to Abelian gauge bosons is a non-commutative
analogue of the usual minimal coupling scheme.

The action for a neutral fermion that couples to an Abelian gauge boson 
in a non-commutative background is
\begin{equation}
S = \int d^4 x \left(\,\overline{\widehat\psi} \star 
i\gamma^\mu\widehat D_\mu \widehat\psi
-m \overline{\widehat\psi} \star \widehat\psi\right).
\label{1}
\end{equation}
Here $\widehat \psi_{\rm L \choose \rm R} = \psi_{\rm L \choose \rm R} +
e \theta^{\nu\rho} A_\rho \pp\nu \psi_{\rm L \choose \rm R}$ and 
$\widehat A_\mu = A_\mu + e\theta^{\rho\nu}A_{\nu}
\left[\partial_{\rho}A_{\mu}-\frac{1}{2}\partial_{\mu}A_{\rho}\right]$ 
is the Abelian NC gauge potential expanded by the Seiberg-Witten map. 
\footnote{Note that instead of Seiberg--Witten map of Dirac fermions $\psi$ 
one can consider a ``chiral'' SW map. 
This SW map is compatible with grand unified models 
where fermion multiplets are chiral \cite{asch}.}

To first order in $\theta$, the gauge-invariant action reads
\begin{eqnarray}
&&S = \int d^4 x \, \left\{ \bar \psi 
\left[i\gamma^\mu \pp\mu  - m\left(1-\frac{e}{2}\theta^{\mu\nu}F_{\mu\nu}\right)\right]\psi
\right. \label{2}\\
&&\left. 
+ ie \theta^{\mu\nu} \left[(\pp\mu \bar \psi) A_\nu \gamma^\rho (\pp\rho \psi)
-  (\pp\rho\bar \psi) A_\nu \gamma^\rho  (\pp\mu  \psi)
+  \bar \psi (\pp\mu A_\rho) \gamma^\rho  (\pp\nu \psi) \right]\right\}.
\nonumber
\end{eqnarray}
Integrating by parts, (\ref{2}) becomes manifestly gauge-invariant and can be 
conveniently expressed by
\begin{eqnarray}
S &=& \int d^4 x \, \bar \psi 
\left[\frac{}{}\left(i\gamma^\mu \pp\mu  - m\right) \,-\,\frac{e}{2}\,F_{\mu\nu}\,\left(
i \,\theta^{\mu\nu\rho}\,\pp\rho -\,\theta^{\mu\nu}\,m\right)\right]\psi
\label{3}\\
&\equiv&\int d^4 x \,  \bar \psi 
\left[\frac{}{}\left(i\gamma^{\mu} \pp\mu  - m\right) 
\right.\nonumber\\  
&&-\left.\frac{e}{2}\theta^{\nu\rho}\left(i\gamma^{\mu}
(F_{\nu\rho}\pp\mu + F_{\mu\nu}\pp\rho+F_{\rho\mu}\pp\nu) - mF_{\nu\rho}\right)\right]\psi,
\nonumber
\end{eqnarray}
where $F_{\mu\nu}=\pp\mu A_{\nu} - \pp\nu A_{\mu}$ and ${\theta}^{\mu\nu\rho}=
{\theta}^{\mu\nu}\gamma^{\rho}+{\theta}^{\nu\rho}\gamma^{\mu}+
{\theta}^{\rho\mu}\gamma^{\nu}$.

The above action presents a tree-level interaction of photons and neutrinos
on non-commutative space-time.
%We could also call it ``the background field anomalous-contact'' interaction.
%Note that in the limit $\theta \rightarrow 0$ Eq. (\ref{3}) becomes standard commutative action.

It is interesting to note that we can write 
\begin{eqnarray}
i\bar \psi F_{\mu\nu}\,\theta^{\mu\nu\rho}\,\pp\rho\psi &=&
F_{\mu\nu}({\theta}^{\mu\nu} T^{\rho}_{\;\:\rho}+{\theta}^{\nu\rho} T^{\mu}_{\;\:\rho}+
{\theta}^{\rho\mu} T^{\nu}_{\;\:\rho})
\nonumber\\
&\equiv& \theta^{\mu\nu}(T^{\rho}_{\;\:\mu}F_{\nu\rho}+T^{\rho}_{\;\:\nu}F_{\rho\mu}+T^{\rho}_{\;\:\rho}F_{\mu\nu})
\label{3a}
\end{eqnarray}
where
\begin{equation}
T^{\mu\nu} = i\bar \psi 
\gamma{^\mu} \partial^{\nu} \psi 
\label{3b}
\end{equation}
represents the stress--energy tensor of commutative 
gauge theory for free fermion fields \cite{AB}.
Hence, for the massless case the Eq. (\ref{3}) 
reduces to the coupling between the stress--energy tensor of the neutrino
$T^{\mu\nu}$ and the symmetric tensor composed from $\theta $ and F.
This nicely illustrates our assertion that we are 
seeing the interaction of the neutrino with a modified photon--$\theta$ background.

%%new:
So far we have not discussed how the terms of the action (\ref{1}) that we have introduced
can be embeded into a model of the full non-commutative electroweak sector. 
We have instead focused
on the interaction term that is relevant for the computation of
the plasmon decay rate. In particular we have not discussed
the form of the gauge kinetic term. Since the choice of model
has some bearing on the resulting phenomenology, in particular 
in the infrared, we shall give a brief overview about the 
various approaches to non-commutative gauge theory. All have in
common that the action resembles Yang-Mills theory, with matrix 
multiplication replaced by $\star$--products.

The most familiar model of non-commutative U(1) is defined in terms 
of Feynman rules that are directly obtained from the action (in momentum space)
without first expanding the $\star$--products or fields in terms of $\theta$. The
resulting phase factors play the role of structure constants in 
ordinary ``commutative'' Yang-Mills theory. The result is that the
beta function resembles that of a non-abelian gauge theory even though
the structure group is abelian~\cite{UV/IR}. The beta function 
with matter in the adjoint has been computed in this approach 
in reference~\cite{Sadooghi:2002jv}, see also~\cite{Nakajima:2002sk}.  
The beta function is negative and we would expect 
problems in the infrared if we were to take this theory
at face value even at low energies.  In particular there
could be condensation of neutrino-antineutrino pairs and
one could question whever it is really justified to 
work in the tree level approximation as we do. 
There is also UV/IR mixing.\footnote{This is not necessarily a bad
thing: UV/IR mixing effects in non-commutative gauge theory on D-branes 
can capture information about the closed string spectrum of the parent
string theory~\cite{Armoni:2003va}.} We are not working with this model, 
but even in this model one can avoid infrared problems in several ways: 
Using reducible representations for the gauge field in the gauge 
kinetic terms of the action, triple gauge couplings, which are responsible for the 
negative beta function, can be eliminated. We could also consider a N=4 supersymmetric 
extension of the model that is softly broken down to N=0~\cite{luis}. 
Finally, infrared problems can be avoided with more sofisticate
quantization and renormalization procedures~\cite{Aigner:2003nd}.
The model has other problems: It is limited to U(N) gauge groups
in the fundamental representation, the fields do not transform
covariantly under coordinate transformations~\cite{Bichl:2001yf}
and there are problems with renormalizability~\cite{Aigner:2003nd}. 

The approach to non-commutative gauge theory that we use belongs to a 
class of models that expand the action in $\theta$ \emph{before}
quantization~\cite{WESS}-\cite{ws1}. Here we do not have any infrared problems, nor do we 
have UV/IR mixing and the beta function is not negative.  
For pure non-commutative Maxwell theory the
photon self-energy has been computed 
to all loop orders in~\cite{Bichl:2001cq}. The beta function is that of ordinary 
abelian gauge theory. 
For neutrinos in the $\star$-adjoint representation
we do not expect any contribution to the beta function
up to the second order in $\theta$,
considering the relevant terms that may enter in the
computation of the beta function at that order.
The computation of higher order corrections to the
beta function in our model is an open project, but
the expected result is a theory without infrared 
problems and in particular without neutrino condensation. 
An objection to the $\theta$-expanded approach is that
it may not capture non-perturbative information
about the non-commutativity of space-time. 
This remains to be seen. 
It  does, however, nicely capture new interactions
induced by spacetime non-commutativity and it can be 
applied to realistic gauge groups like U(1)$\times$SU(2) in 
the present case.

The model is meant to provide an effective description of space-time non-commutativity
involving the photon--neutrino contact interaction. 
Therefore, we treat our action as an effective action, disregarding renormalizability
in the ordinary sense. 
This approach is similar to chiral dynamics in pion physics. 
As we have discussed above, it differes fundamentally from other approaches
based on $\star$--products that are not $\theta$-expanded and do not use
the Seiberg-Witten map: We expand the action up to a certain fixed order in $\theta$
\emph{before} quantization.
The effective theory obtained appears to be anomaly free \cite{mart}.

Concerning the physics considered, the picture
that we have in mind is that of a space-time that has a continuous 
`commutative' description at low energies and long distances, but a non-commutative
structure at high energies and short distances. There could be some kind of phase-transition
involved. At high energies we can model space-time using $\star$--products.
This description is not valid at low energies. 
On the technical side this means that by expanding up to a certain order in $\theta$
and considering renormalization of this truncated theory up to the same order in $\theta$
there will not arrise any infra red problem. 
This reflects very well our assumption: 
At low energies and large distances the non-commutative theory has to be modified.

We now apply our model to the decay of plasmons
into neutrino - anti-neutrino pairs induced by a hypothetical stellar non-commutative
space-time structure. The resulting neutrinos can escape from the star and thereby lead to an energy loss. 
To obtain the ``transverse plasmon'' decay rate in stars on 
the scale of non-commutativity, we start
with the action determining the $\gamma\nu\bar\nu$ interaction.
From Eq. (\ref{2}) we extract, for left or right and possibly massive neutrinos, 
the following Feynman rule for the gauge invariant 
${\gamma}(q)\to {\nu}(k'){\bar {\nu}}(k)$ vertex in momentum space:
\begin{equation}
{\Gamma}^{\mu}_{\rm L \choose \rm R}({\nu}{\bar {\nu}}{\gamma})
=ie\frac{1}{2}(1 \mp \gamma_5)
\left[(q\theta k)\gamma^{\mu}+({\not \!k}-m_{\nu}){\widetilde q}^{\mu}-
{\not \!q}{\widetilde k}^{\mu}\right].
\label{4}
\end{equation}
Here we have used the notation ${\widetilde q}^{\mu} \equiv {\theta^{\mu\nu}q_{\nu}}$,
${\widetilde k}^{\mu} \equiv {\theta^{\mu\nu}k_{\nu}}$.
In the case of massless neutrinos, the vertex (\ref{4}) becomes symmetric:
\begin{eqnarray}
{\Gamma}^{\mu}_{\rm L \choose \rm R}({\nu}{\bar {\nu}}{\gamma})
=ie\frac{1}{2}(1 \mp \gamma_5){\theta}^{\mu\nu\rho}k_{\nu}q_{\rho}.
\label{5}
\end{eqnarray}

In stellar plasma, the dispersion relation of photons is identical with that of 
a massive particle \cite{Jancovici}--\cite{sal}:
\begin{equation}
q^2 \equiv {\rm E}_\gamma^2-{\bf q}_\gamma^2=\omega_{\rm pl}^2
\end{equation}
with $\omega_{\rm pl}$ being the plasma frequency. 

From the gauge-invariant amplitude 
${\cal M}_{\gamma {\nu} {\bar{\nu}}}$ in momentum space 
for the plasmon (off-shell photon) decay to the
left and/or right massive neutrinos in our model, we have 
\footnote{Note that this result is independent on different choices of the Seiberg--Witten map
for righthanded Dirac fermions, see footenote 2.}
\begin{eqnarray}
\sum_{\rm pol.} |{\cal M}_{\gamma {\nu} {\bar{\nu}}}|^2 =
 4e^2\left[\left(q^2 -2m_{\nu}^2\right)
 \left(m_{\nu}^2{\widetilde q}^2-(q\theta k)^2\right) 
 +m_{\nu}^2 q^2 ({\widetilde k}^2-{\widetilde k}{\widetilde q})\right].
\nonumber
\end{eqnarray}
Phase-space integration of this expression then gives
\begin{eqnarray}
&&\Gamma(\gamma_{\rm pl}\rightarrow {\bar{\nu}}_{\rm L\choose \rm R}\nu_{\rm L\choose \rm R})
= \frac{\alpha}{48}\frac{\omega^6_{\rm pl}}{{\rm E}_{\gamma}\Lambda^4_{\rm NC}}
\sqrt{1-4\frac{m_{\nu}^2}{\omega_{\rm pl}^2}}\\
&&\times \left[\left(1+2\frac{m^2_{\nu}}{\omega^2_{\rm pl}}
-12\frac{m^4_{\nu}}{\omega^4_{\rm pl}}\right)
\sum_{i=1}^{3}(c^{0i})^2 + 
2\frac{m^2_{\nu}}{\omega^2_{\rm pl}}
\left(1-4\frac{m^2_{\nu}}{\omega^2_{\rm pl}}\right)
\sum_{{i,j=1 \atop i<j}}^{3}(c^{ij})^2\right].
\nonumber
\end{eqnarray}

In the above formula we have
parametrized the $c_{0i}$'s by introducing the angles characterizing 
the background ${\theta}^{\mu\nu}$ field of the theory \cite{HK}:
\begin{eqnarray}
c_{01}=\cos\xi,\;\;c_{02}=\sin\xi \;\cos\zeta,\;\;c_{03}=\sin\xi \;\sin\zeta,
\nonumber
\end{eqnarray}
where $\xi$ is the angle between the ${\vec E}_{\theta}$ field and 
the direction of the incident beam,
i.e. the photon axes. The angle $\zeta$ defines the origin of the $\phi$ axis.  
The $c_{0i}$'s are not independent; in pulling out the overall scale 
$\Lambda_{\rm NC}$, we can always
impose the constraint ${\vec E}^2_{\theta}\equiv \sum_{i=1}^3 (c^{0i})^2=1$. 
Here we consider three physical cases:
$\xi=0,\;\pi/4,\;\pi/2$, which for $\zeta = \pi/2$ satisfy the imposed constraint.
This parametrization provides a good physical interpretation of the NC effects \cite{HK}.

In the rest frame of the medium, the decay rate of a ``transverse plasmon'',
of energy ${\rm E}_\gamma$ for the left--left and/or right--right 
massless neutrinos and for the constraint 
${\vec E}^2_{\theta}=1$, is given by
\begin{equation}
\Gamma_{\rm NC}
(\gamma_{\rm pl}\to \nu_{\rm L\choose \rm R}\bar\nu_{\rm L\choose \rm R})
=\frac{\alpha}{48}\,\frac{1}{\Lambda_{\rm NC}^4}\,
\frac{\omega_{\rm pl}^6}{{\rm E}_\gamma}\,.
\end{equation}

The Standard Model (SM) photon--neutrino interaction at tree
level does not exist. However, the effective photon--neutrino--neutrino vertex 
$\Gamma^{\mu}_{\rm eff}(\gamma\nu\bar\nu)$ is
generated through 1-loop diagrams, which are very well known in heavy-quark physics as ``penguin
diagrams''. Such effective interactions give non-zero charge radius, as well as the
contribution to the ``transverse plasmon'' decay rate \cite{Bernstein}--\cite{BPV}. 
For details, see Ref. \cite{as}. Finally, note that the dipole moment operator
$\sim em_{\nu}G_{\rm F}{\bar\psi}\sigma_{\mu\nu}\psi F^{\mu\nu}$, 
also generated by the ``neutrino-penguin diagram'', 
gives very small contributions because of the smallness of the neutrino mass, 
i.e. $m_{\nu} < 1$ eV \cite{nobel}.

The corresponding SM neutrino-penguin-loop result for the ``transverse plasmon'' decay rate is \cite{as}:
\begin{equation}
\Gamma_{\rm SM}\left(\gamma_{\rm pl}\to {\nu_{\rm L}}{\bar\nu}_{\rm L}\right)
=\frac{{\rm c}_{\rm v}^2 G_{\rm F}^2}{48\pi^2 \alpha}\;\frac{\omega_{\rm pl}^6}{{\rm E}_\gamma}.
\end{equation}
For $\nu_e$, we have ${\rm c}_{\rm v}=\frac{1}{2}+2\sin^2\Theta_{\rm W}$,
while for $\nu_\mu$ and $\nu_\tau$ we have
${\rm c}_{\rm v}=-\frac{1}{2}+2\sin^2\Theta_{\rm W}$.  Comparing the rate
of the decays into all three neutrino families, we thus need
to include a factor of~3 for the NC result, while 
${\rm c}_{\rm v}^2 = 0.79$ for the SM result \cite{rpp}. From the ratio of the rates 
\begin{equation}
\Re\equiv\frac{\sum_{\rm flavours}
\Gamma_{\rm NC}
\left(\gamma_{\rm pl}\to {\nu_{\rm L}}{\bar\nu}_{\rm L} + {\nu_{\rm R}}{\bar\nu}_{\rm R}\right)}
{\sum_{\rm flavours}\Gamma_{\rm SM}(\gamma_{\rm pl}\to {\nu_{\rm L}}{\bar\nu}_{\rm L})}
=\frac{6\pi^2\alpha^2}{{\rm c}_{\rm v}^2 G_{\rm F}^2\Lambda_{\rm NC}^4},
\end{equation}
we obtain
\begin{equation}
\Lambda_{\rm NC}\;= \;\frac{80.8}{\Re^{1/4}} \;(\rm GeV).
\end{equation}
A standard argument involving globular cluster stars tells us that any
new energy loss mechanism must not excessively exceed the standard neutrino losses; 
see section 3.1 in Ref. \cite{raffelt}.
Expressed in another way, we should approximately
require $\Re<1$, translating into
\begin{equation}
\Lambda_{\rm NC}>\left(\frac{6\pi^2\alpha^2}{{\rm c}_{\rm v}^2 G_{\rm F}^2}\right)^{1/4}
\cong 81~{\rm GeV}\,.
\end{equation}
If sterile neutrinos ($\nu_{\rm R}$) do not exist,
the scale of non-commutativity is approximately $\Lambda_{\rm NC}> 68~{\rm GeV}$.

The advantage of our approach to 
the anomalous $\gamma\nu\bar\nu$ interaction,
via non-commutative Abelian gauge field theory, lies in the fact that,
contrary to the SM approach, 
photons are also coupled to the sterile neutrinos in the same, 
U(1)-gauge-invariant, way as the left-handed ones. 
The electromagnetic gauge invariance of the $\gamma\nu\bar\nu$ amplitude 
comes automatically, since the starting action is manifestly 
U(1)-gauge-invariant. The interaction (\ref{2}) produces
extra contributions relative to the SM in the non-commutative background.

The non-commutativity scale depends on the requirement $\Re<1$ and from this aspect,
the constraint $\Lambda_{\rm NC}> 80~{\rm GeV}$,
obtained from the energy loss in globular stellar clusters,
represents the lower bound on the scale of non-commutative gauge field theories.
\footnote{ Note that, for example, $\Lambda_{\rm NC}>144$ GeV for $\Re<1/10$.}
It also depends on the strength of the non-commutative coupling constant $\kappa$
which we have taken to be $\kappa=1$. 

Compared with other bounds, see \cite{HK},
the bound that we have obtained is relatively low. 
However, it is based on a completely new interaction channel and a completely different 
``laboratory'' than other constraints and as such appears worth communicating.
\vspace{0.8cm}

We would like to thank  
L. Alvarez-Gaume, A. Armoni, P. Aschieri,
P. Minkowski, K. Passek-Kumeri\v cki, Leo Stodolsky and R. Wulkenhaar 
for stimulating and helpful discussions.

The work of J.T. is supported by the Ministry of Science and Technology 
of Croatia under Contract No. 0098002. G.R. acknowledges partial support by the 
Deutsche Forschungsgemeinschaft under grant No. SFB 375.

\end{document}